\documentclass[aps,pra,twocolumn]{revtex4-1}
\usepackage{graphicx}
\usepackage{amssymb,amsfonts,amsmath}
\usepackage[colorlinks=true,citecolor=Cerulean,linkcolor=RubineRed,urlcolor=Cerulean]{hyperref}
\hypersetup{breaklinks=true}
\usepackage{graphicx}
\usepackage{color}
\usepackage[usenames,dvipsnames]{xcolor}
\usepackage{epstopdf}
\usepackage{units}

\begin{document}

\title{Fractals on a benchtop: Observing fractal dimension in a resistor network}

\author{C.E.~Creffield}
\affiliation{Departamento de F\'isica de Materiales, Universidad
Complutense de Madrid, E-28040 Madrid, Spain}

\date{\today}

\begin{abstract}
Our first experience of dimension typically comes in the intuitive
Euclidean sense: a line is one dimensional, a plane is
two-dimensional, and a volume is three-dimensional. However, following
the work of Mandelbrot \cite{mandelbrot}, systems with a fractional
dimension, ``fractals'', now play an important
role in science. The novelty of encountering fractional dimension,
and the intrinsic beauty of many fractals, have a strong appeal to students and
provide a powerful teaching tool. I describe here a low-cost and convenient 
experimental method for observing fractal dimension, by measuring the power-law scaling of
the resistance of a fractal network of resistors. The experiments are quick to perform,
and the students enjoy both the construction of the network and the
collaboration required to create the largest networks. Learning outcomes include
analysis of resistor networks beyond the elementary series and parallel combinations,
scaling laws, and an introduction to fractional dimension.
\end{abstract}

\maketitle

Examples of fractals from the natural world cover an astonishing range
and variety. The coastline of a country, for example,
retains the same general pattern of jaggedness
over huge range of length scales, and is described
by a fractal dimension between one and two. 
Analogously, the self-similar structures of ferns and broccoli,
and the forms of the respiratory and circulatory systems,
can be described as fractals.
Fractals also appear in quantum systems, such as 
``Hofstadter's butterfly'' which describes the conductance of a
two-dimensional system under a magnetic field \cite{hofstadter}, and,
more recently, the quantum transport of electrons though a fractal
nanostructure \cite{cristiane}.

\begin{figure}
\begin{center}
\includegraphics[width=0.40\textwidth,clip=true]{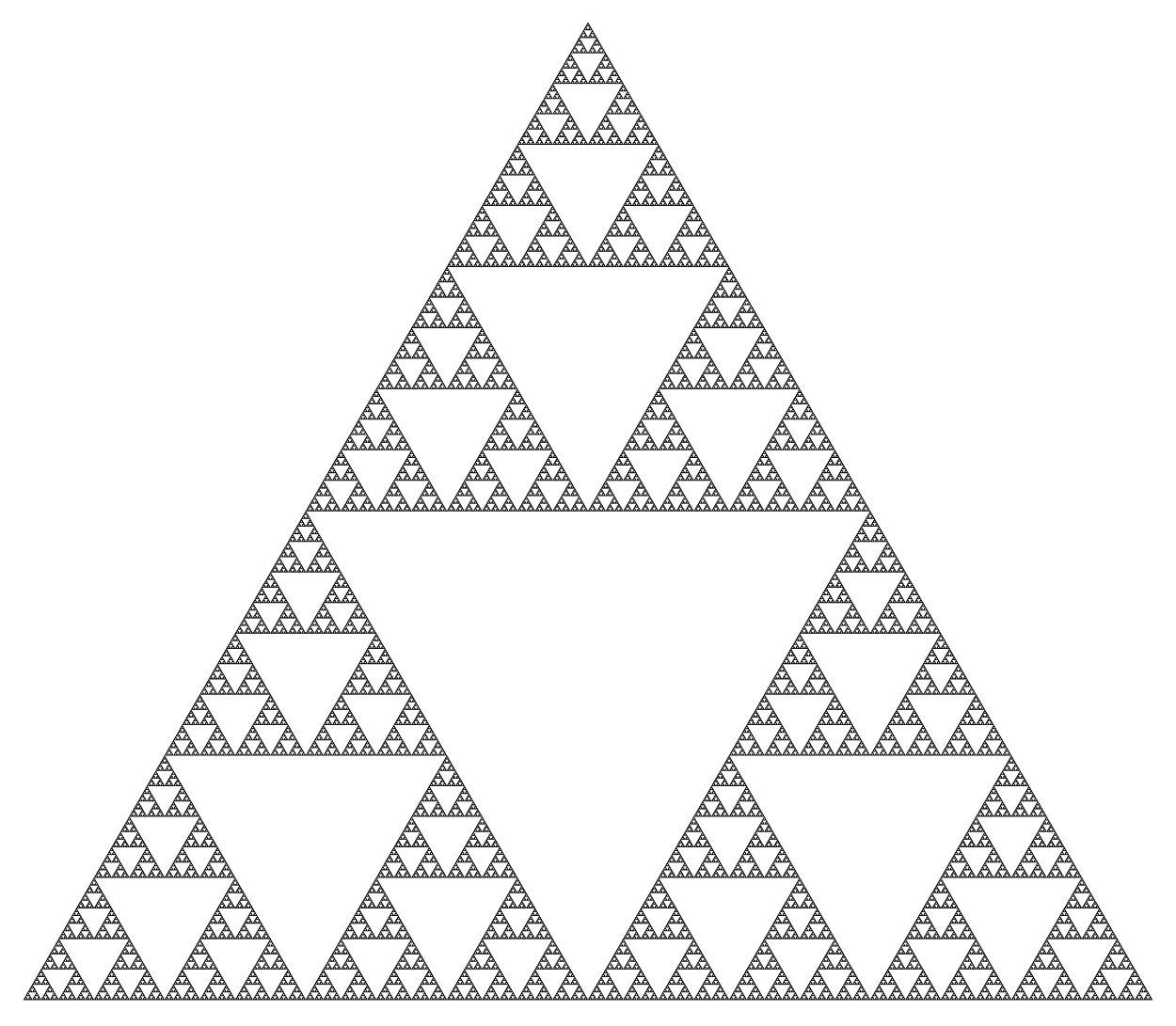}
\end{center}
\caption{The fractal structure known as Sierpinski's gasket has the overall
shape of an equilateral triangle, subdivided recursively into smaller
equilateral triangles. It is self-similar at all length scales, 
so no matter how much one enlarges a small area of the gasket, one encounters the same
pattern of recursively nested triangles.}
\label{ideal}
\end{figure}

Despite their ubiquity, however, it is unusual for students to encounter fractals
in their undergraduate studies, especially in the laboratory. 
Some pioneering efforts in this direction
include the measurement of the fractal dimension of crumpled paper balls
\cite{gomes_balls,ko_balls}, the self-similar structure of breads \cite{bread},
and measuring the fractal dimension of cauliflower \cite{cauliflower}.
Although these experiments indeed yield estimates of the fractional dimension of
these objects, a slightly unsatisfactory aspect is the absence of well-defined
theoretical estimates to compare with the results. In this note I 
describe a method to measure the power-law scaling of the resistance of a
fractal network known as Sierpinski's gasket (see Fig. \ref{ideal}), 
which can be computed exactly.
By measuring the resistance of several system sizes, students are able
to confirm the power-law behaviour of the resistance, and
extract the fractal dimension with high precision.

\begin{figure}
\begin{center}
\includegraphics[width=0.45\textwidth,angle=0,clip=true]{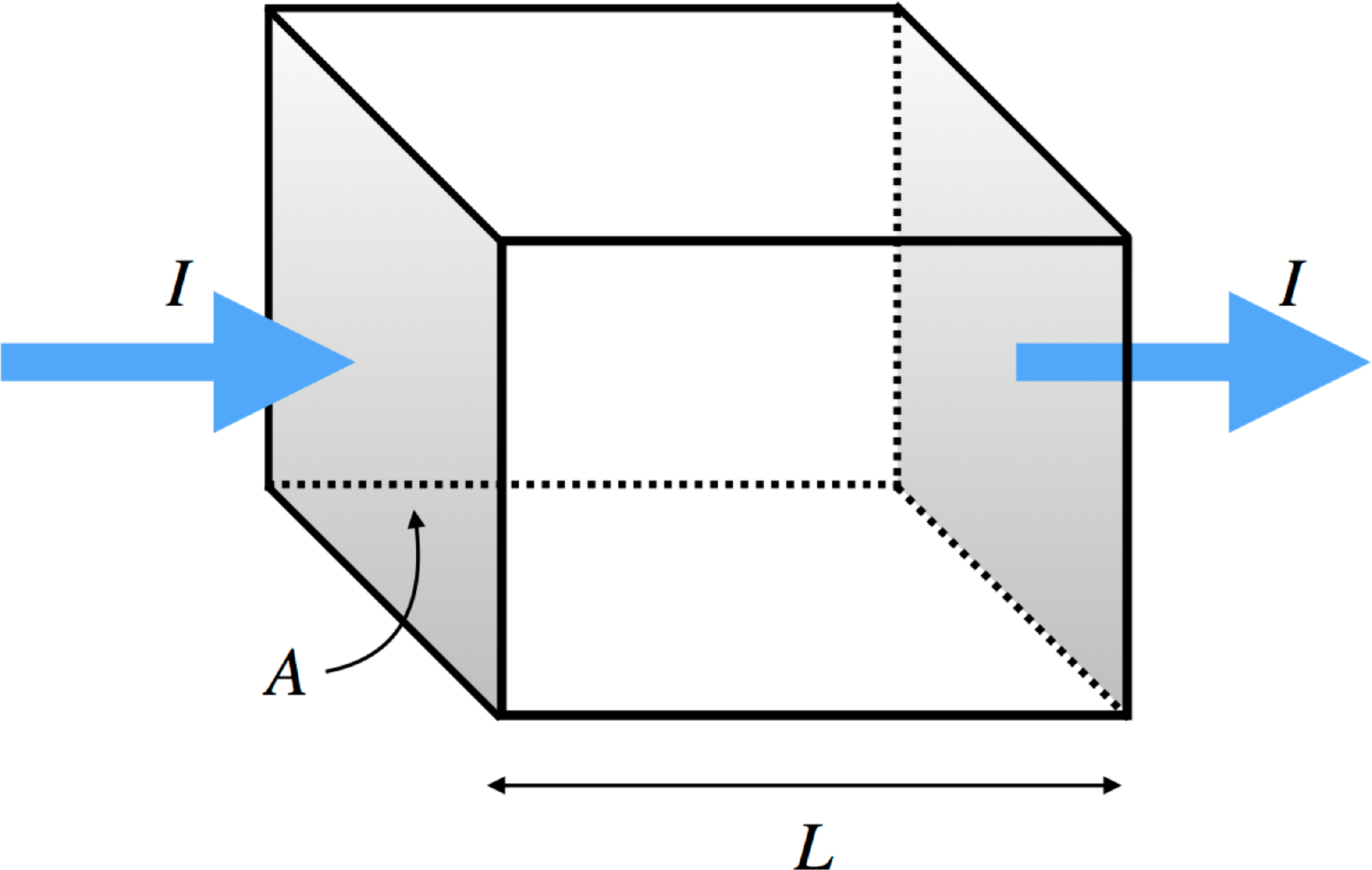}
\end{center}
\caption{Geometry to define the resistance of a block of material
with resistivity $\rho$. A current $I$ enters and leaves through the
shaded faces which have each a cross-sectional area of $A$. If the length
of the block is $L$, its resistance is $R = \rho L / A$.}
\label{euclidian}
\end{figure}

Let us first see how the dimension of a system influences its resistance.
Fig. \ref{euclidian} shows a standard set-up; a rectangular block of a material with
resistivity
$\rho$, that has length $L$ and cross-sectional area $A$, admits a 
current of $I$ when a potential difference is applied across its length. 
The resistance of this block is simply given by $R = \rho L/A$. Now let
us consider making the cross-section very small $A = d^2$ where $d \ll L$.
This produces an effectively one-dimensional conductor, and
clearly its resistance will scale {\em linearly} with its length,
$R \propto L$. Now let us change one of the transverse dimensions to
$L$, giving a cross-sectional area of $A = L d$,
so that the sample forms a thin sheet.
In this case we obtain the somewhat counter-intuitive result that $R$ is constant,
that is, the resistance of a two-dimensional conductor does {\em not} depend on its
area. Finally, if we consider a cube by setting $A = L^2$, it is clear that 
$R \propto 1/L$. If we write the scaling of the resistance 
as $R \propto L^\eta$, we can combine these results to show that the 
Euclidean dimension of the system, $D_E$, is related to the scaling of the
resistance by the relation:
\begin{equation}
\eta = 2 - D_E \ .
\label{scaling}
\end{equation}

\begin{figure}
\begin{center}
\includegraphics[width=0.32\textwidth,angle=-90,clip=true]{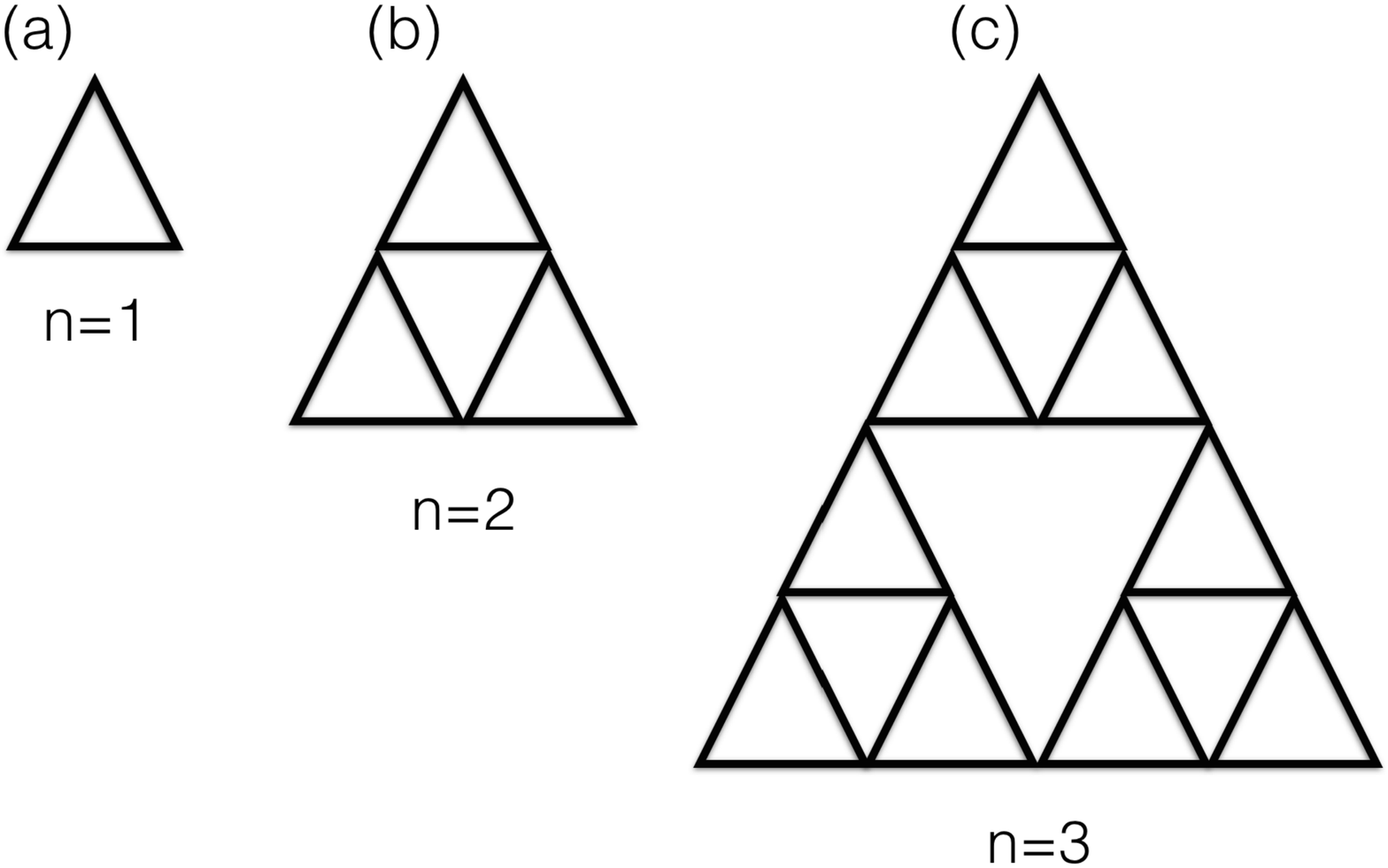}
\end{center}
\caption{The fractal Sierpinski gasket shown in Fig.\ref{ideal}
can be obtained as the
$n \rightarrow \infty$ limit of recursively dividing a triangle.
Here we show the (a) $n=1$, (b) $n=2$, and (c) $n=3$ levels of
this process. Compare with the resistor networks shown in
Fig. \ref{resistors}  which provide a physical implementation 
of these structures.}
\label{triangle}
\end{figure}

{\em Fractional dimension -- }
Looking at Fig. \ref{ideal}, the Sierpinski gasket intuitively has a dimension 
that is smaller
than that of a standard two-dimensional triangle, but is larger than
a line. How can we define such a fractional dimension?
The key is to study how the size, or ``hypervolume'', of an object
increases as it is enlarged. 
For an object with integer dimension $D$, doubling its
side-length increases its hypervolume by a factor of $2^D$. For example,
doubling the side-length of a square produces a new square that is 4 times larger
than the original indicating that (as expected) $D = 2$,
while doubling the side-length of a cube increases its volume by
a factor of 8, meaning than $D = 3$. This permits an alternative definition 
of dimension, 
the Hausdorff dimension, which agrees with $D_E$ for integer dimensions,
but can also be applied to fractal objects. From Fig. \ref{triangle}
it is clear that doubling the length of the triangle sides produces three 
copies of the Sierpinski gasket; for example the $n=3$ level network has twice
 the side-length of the $n=2$ network, and contains three copies of it.
Solving the relation $3 = 2^D$ reveals that the
Hausdorff dimension of the Sierpinski gasket is $D = \log_2 3 \simeq 1.585$.
As anticipated, this value lies between 1 and 2.

{\em Resistance scaling -- }
It is tempting to obtain the scaling of the resistance of Sierpinski's gasket,
measured between two of its exterior corners, 
by directly substituting this value for $D$ in Eq. \ref{scaling}.
The fractal nature of the system, however, means that we must calculate
its resistance with more care. 
Fig. \ref{triangle} shows how the gasket can be produced by a recursive process.
Level 1 (Fig. \ref{triangle}a) consists of a single upward-pointing triangle. 
To reach level 2 (Fig. \ref{triangle}b) this triangle is enlarged by a factor 
of two, and then subdivided to form three upward-pointing triangles.
The same procedure -- scaling every upward-pointing triangle by two and then
subdividing into three -- is then applied to create the level 3 network
(Fig. \ref{triangle}c). Continuing this process indefinitely
produces the true Sierpinski gasket shown in Fig. \ref{ideal}.

\begin{figure}
\begin{center}
\includegraphics[width=0.32\textwidth,angle=-90,clip=true]{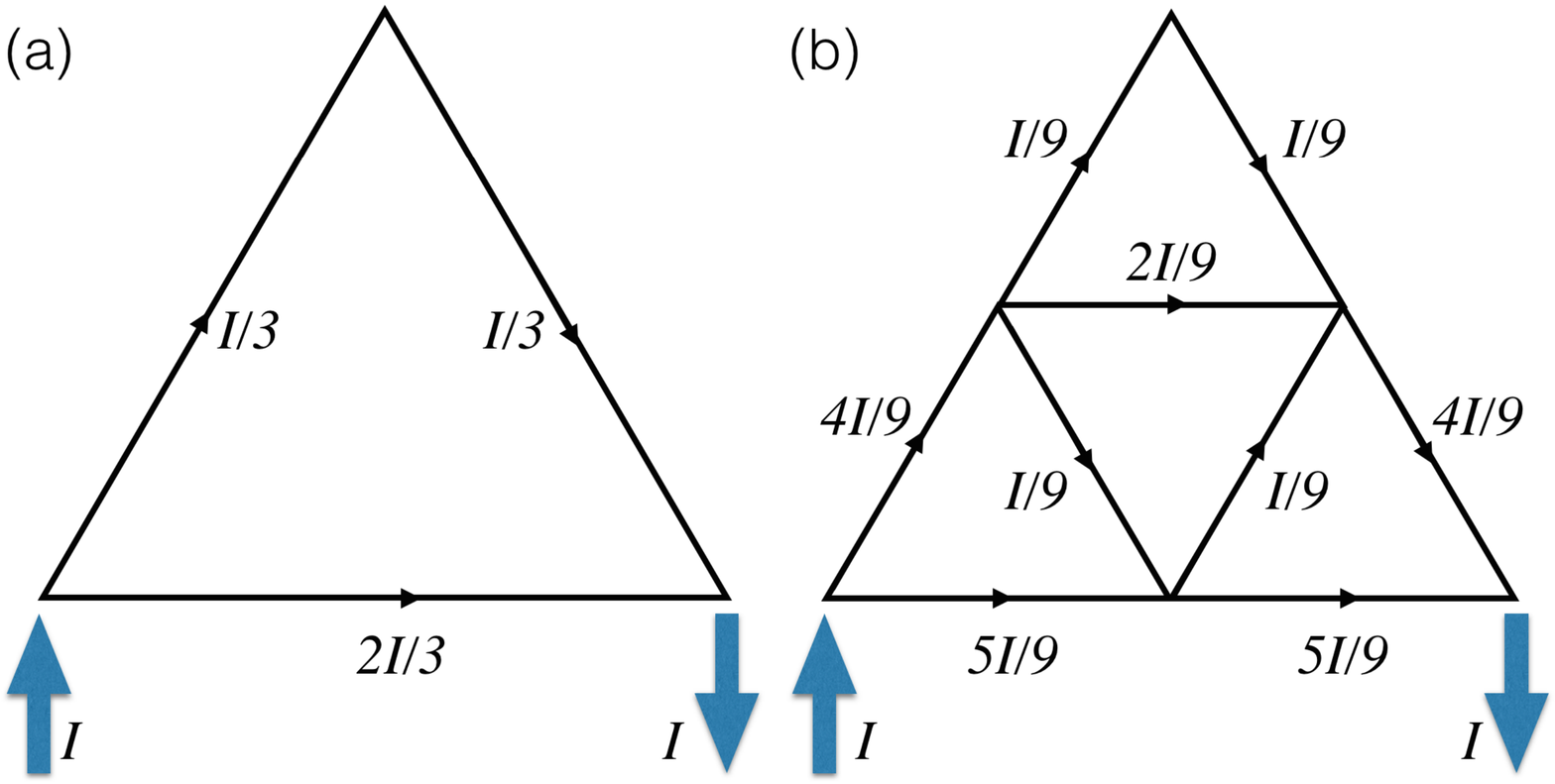}
\end{center}
\caption{Current flowing through the (a) level-1 and (b) level-2 
approximations to Sierpinski's gasket. Kirchoff's laws are
used to find the currents flowing through each resistance.} 
\label{kirchoff}
\end{figure}

The resistance of the level-1 configuration can be obtained easily by
expressing it in terms of series and parallel resistances. Higher levels 
are more challenging to treat in this way, 
but can be analyzed
(see Ref. \cite{teaching_resistances})
using the ``star-triangle'' method (also sometimes called $\Delta-Y$ method),
or applying nodal analysis \cite{platonic}.
However, an alternative approach is to use Kirchoff's laws.
The level-2 system shown in Fig. \ref{kirchoff}b is sufficiently simple 
that the internal currents  
can be obtained by the students (with some prompting) through trial and error, 
noting that
the currents entering a node must equal the currents exiting it, and that
the sum of the currents around a closed loop must be equal to zero.
Alternatively it is straightforward to write the simultaneous equations
satisfied by the currents,
and solve the problem algebraically. 
We can see directly that the majority of the current flows along the bottom
edge, indicating that the current flow is
predominantly one-dimensional, with the incursions into the bulk of the
system producing small modifications from this behaviour. 
Let us suppose that each side of every triangle has a resistance of $R_0$.
From the configurations of current shown in Fig. \ref{kirchoff} it can be seen that
the net resistance between the two corners of the level-1 system 
is given by $2 R_0 / 3$, while the level-2 resistance is $10 R_0/9$,
the two differing by a factor of $5/3$.
That is to say, for the purposes of measuring  resistance by connecting
external probes, the level-2 network composed of resistors of value $R$
can be replaced by a level-1 network with
resistors of $5 R/3$.

This process can be repeated to analyze ever-higher fractal levels.
A level-3 network (see Fig. \ref{triangle}c) can
first be reduced to a level-2 network by scaling the resistances by a factor
of $5/3$, which, as seen above, can then be reduced in turn to an level-1
network, introducing another factor of $5/3$. The resistance of
a level-$n$ network is thus given by the power-law
\begin{equation}
R_n = \frac{2 R_0}{3} \  \left( \frac{5}{3} \right)^{n-1} \ .
\label{resistance}
\end{equation}
This form of analysis, finding a scaling law for the value
of a quantity as fine structure is successively eliminated, is very similar
to the real-space renormalization group transformations used to analyze
second-order phase transitions \cite{kadanoff}.

From this scaling analysis we now know the corner-corner resistance
of a level-$n$ approximation to Sierpinski's gasket.
It is also clear (see Fig. \ref{triangle}) that the side-length
of a level-$n$ network is given by $L = 2^{ n - 1 }$. 
This relationship can be used to 
eliminate $\left( n-1 \right)$ from Eq. \ref{resistance}, in order to obtain the
dependence of the resistance on $L$. Some straightforward algebra
yields the result 
\begin{equation}
R(L) = \frac{2 R_0}{3} \ L^{\log_2  \left( 5 / 3 \right)} 
\simeq \frac{2 R_0}{3} \ L^{0.737} \ . 
\label{power}
\end{equation}
If, as before,
we write the scaling of the resistance as $R \propto L^\eta$, we can
thus obtain the exponent, $\eta$, as:
\begin{eqnarray}
\eta = &=& \log_2 \left( \frac{5}{3} \right) \nonumber \\
{ }    &=& \log_2 5 - D
\end{eqnarray}
where $D = \log_2 3$ is the Hausdorff dimension of the Sierpinski gasket.
This equation is the fractional dimension analogue to Eq. \ref{scaling}.

\begin{figure}
\begin{center}
\includegraphics[width=0.32\textwidth,angle=-90,clip=true]{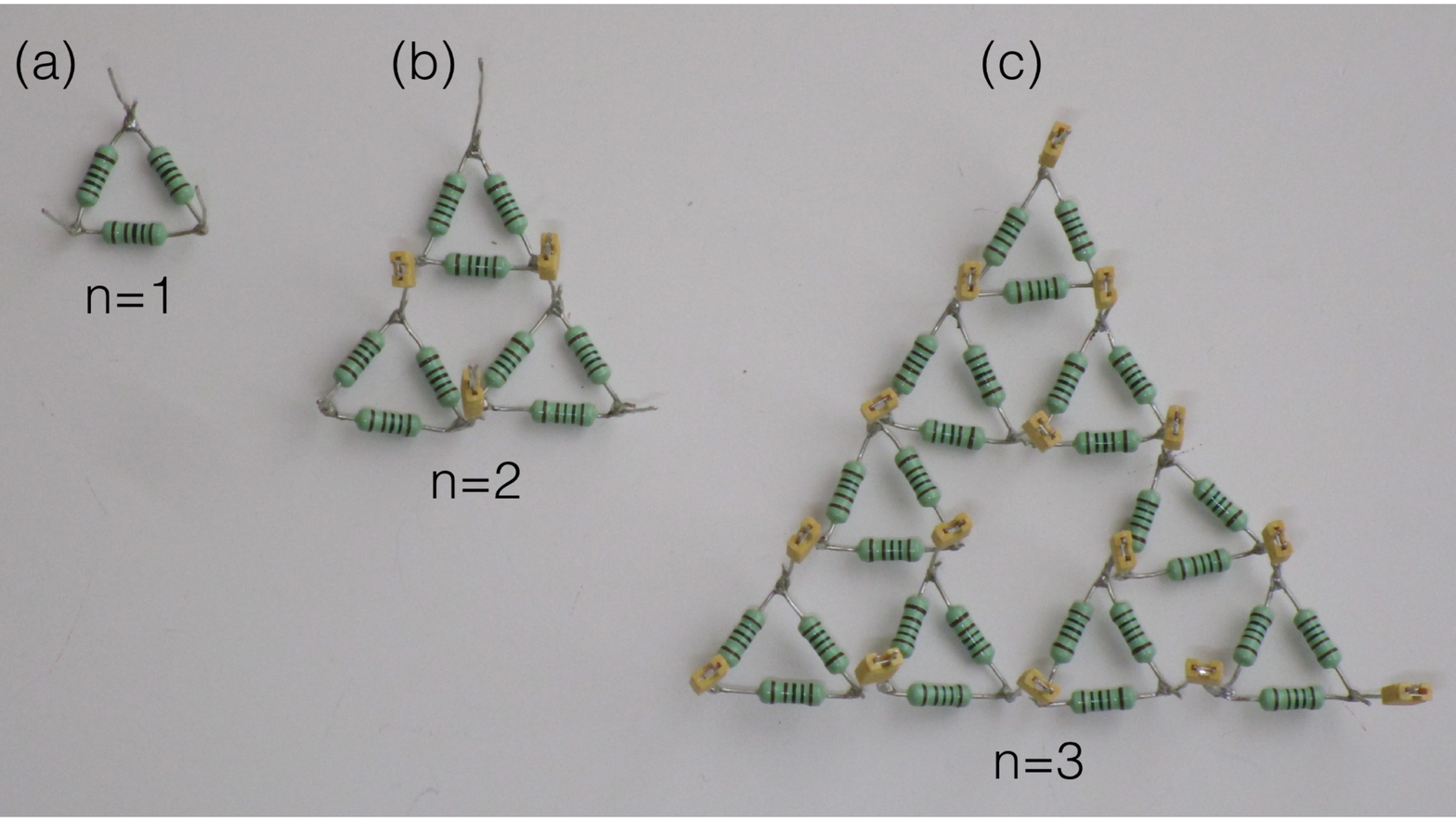}
\end{center}
\caption{(a) The basic element of the resistor network, three
1 k$\Omega$ resistors connected in a triangle.
(b) Level-2 of the Sierpinski gasket, obtained by
connecting three triangular elements with jumper links.
(c) Level-3 of the Sierpinski gasket.}
\label{resistors}
\end{figure}

\begin{figure}
\begin{center}
\includegraphics[width=0.35\textwidth,angle=-90,clip=true]{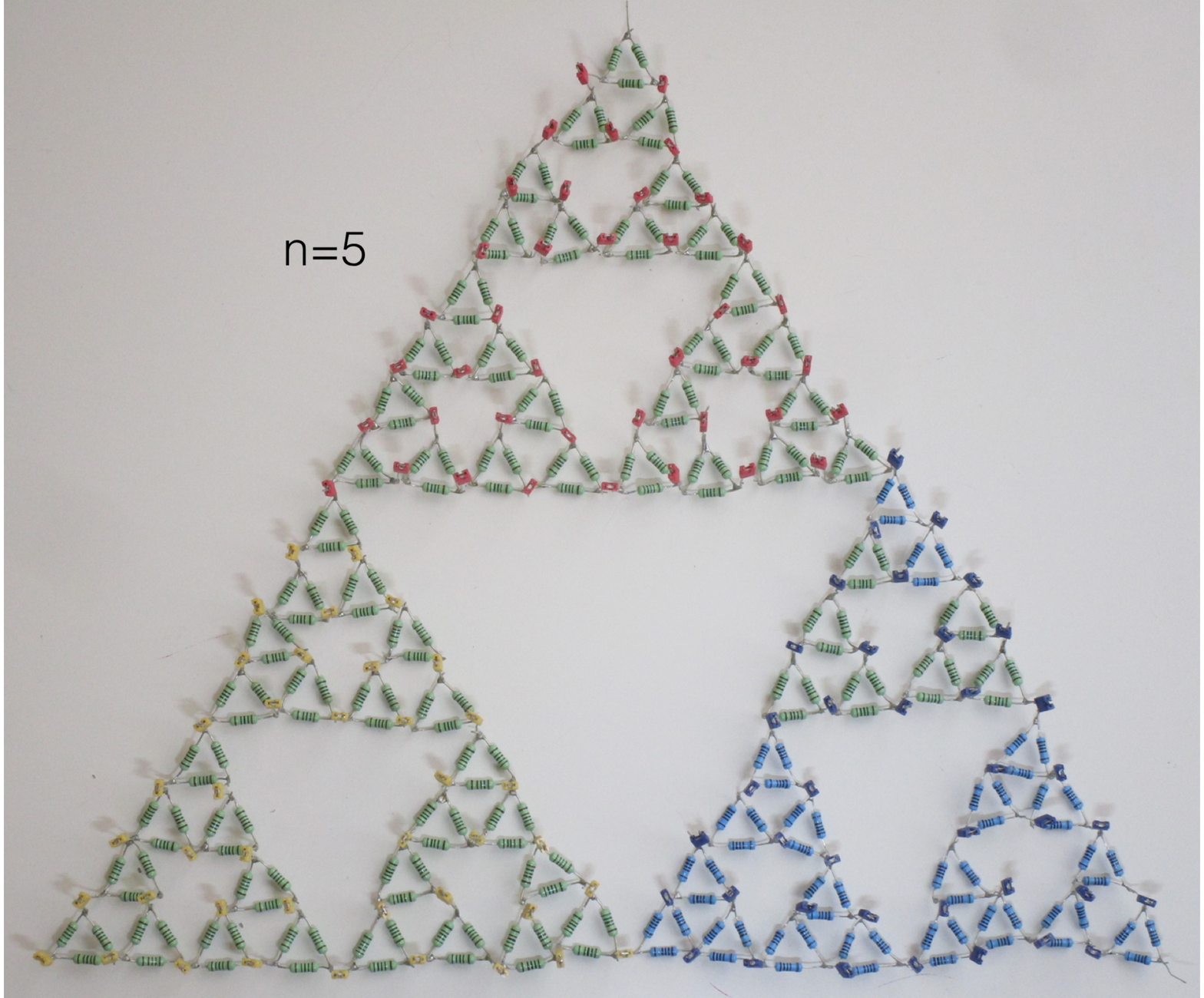}
\end{center}
\caption{The highest level ($n=5$) network considered
in this work, containing 81 triangular elements and 243 individual resistors.}
\label{resistor_gasket}
\end{figure}

{\em Implementation -- }
To provide a physical realization of the networks shown in Fig. \ref{triangle},
the basic triangular building block was obtained
by soldering three 1 k$\Omega$
metal film resistors (0.5W, tolerance 1\%) in a triangle as
shown in Fig. \ref{resistors}a. These blocks could then be connected with
jumper links, cheap and easily available components used in microelectronics,
to construct the various resistor networks.
The corner-corner resistance can be easily measured using an ohmmeter,
or alternatively by connecting the network across a bench power supply
and measuring the input current and voltage dropped across it.
Having connected three triangular elements to create the level-2
network, students can then build and connect further copies to create
the level-3 network, and so on. The largest network used in practice was
of level 5 (Fig. \ref{resistor_gasket}), for which three groups of students 
can contribute their
level-4 networks. In principle even higher orders can be obtained, if
sufficient triangular elements are prepared.

\begin{figure}
\begin{center}
\includegraphics[width=0.5\textwidth,clip=true]{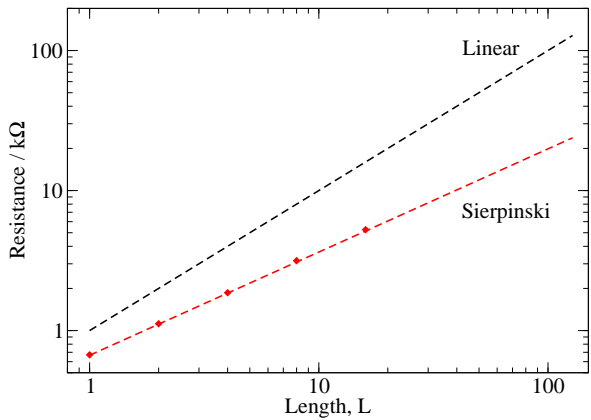}
\end{center}
\caption{Power-law scaling of the corner-corner resistance in
a Sierpinski resistance network. The error bars on the data points are
smaller than the symbol size. The resistance follows a power-law dependence
on the side-length of the network, following the theoretical
result (Eq. \ref{theory}, shown by the dashed red line) 
to an excellent degree of accuracy.
For comparison, the linear scaling of a one-dimensional array of
resistors is shown by the black dashed line.}
\label{data}
\end{figure}

To observe the fractional scaling of the resistance, we measure
the corner-corner resistance for the different fractal levels.
From Eq. \ref{power} it is clear that 
\begin{equation}
\log R = \eta \log L + \log \left( 2 R_0 / 3 \right) \ ,
\label{theory}
\end{equation}
where $\eta = \log_2 \left( 5 / 3 \right)$, and so making a log-log
plot of this data will yield a straight line with a slope of $\eta$.
Fig. \ref{data} shows some typical data obtained in this experiment.
Error bars were estimated from the statistical variance of
the three resistances measured between the pairs
of exterior corners of each network. Making a linear regression to these
data points yields a value of $\eta = 0.743 \pm 0.002$, in
excellent agreement with the theoretical value of $\eta = 0.737$. 
Students found it particularly gratifying to obtain results of
such high precision, using such a ``low-tech'' construction method.

The students clearly enjoyed the process of building the resistor networks,
and the process allowed them to become more engaged with the
apparatus than normal. In electricity and magnetism practicals, the majority
of the components are provided essentially as ``black boxes'' to be
plugged into circuit boards in highly specific arrangements, and so the
construction process proved to be a stimulating change. In particular
students exhibited considerable pride in successfully completing
the largest (level-5) network which has a rather striking, and notably fractal,
appearance when laid out on the bench.

\acknowledgments
I would like to thank Alan L. Smith
for inspiring this investigation.
This work was supported by Spain's MINECO through
grant FIS2017-84368-P.

\bibliographystyle{aipnum4-1}
\bibliography{fractal_bib}

\end{document}